%% file: paper.tex
\documentclass[conference]{IEEEtran}
\usepackage{color}     
\usepackage{epsf,psfrag,amssymb,amsfonts,latexsym,cite,verbatim,enumerate,subfigure}
\usepackage{srcltx,amsmath,cases, graphicx}
\usepackage{times, multirow, array}
\usepackage[mathscr]{eucal}

\input macros

\input labelfig.tex

\def\scalefig#1{\epsfxsize #1\textwidth}

\newcommand {\Ebb}{{\mathbb{E}}}

\IEEEoverridecommandlockouts 

\title{\LARGE {Pilot Beam Sequence Design for Channel Estimation in Millimeter-Wave MIMO Systems: A POMDP Framework}}

\author{
 Junyeong Seo, {\em Student~Member, IEEE}, Youngchul
Sung, {\em
Senior~Member, IEEE} \\ Gilwon Lee, and Donggun Kim, {\em Student~Members, IEEE} \\
\thanks{The authors are with the Dept. of Electrical Engineering,  KAIST, Daejeon 305-701, South
Korea. E-mail:\{jyseo@, ysung@ee., gwlee@, and dg.kim@\}kaist.ac.kr.
This work was supported by ICT R\&D program of MSIP/IITP [ 11-911-04-001, Development of Adaptive Beam Multiple Access Technology  without Interference based on Antenna Node Grouping].  }
}

\begin{document}

\maketitle

\begin{abstract}
    In this paper, adaptive pilot beam sequence design for channel estimation
in large millimeter-wave (mmWave) MIMO systems is considered.  By
exploiting the sparsity of
    mmWave MIMO channels with the
virtual channel representation and imposing a Markovian random
walk assumption on the physical movement of the line-of-sight
(LOS) and reflection clusters, it is shown that the sparse channel
estimation problem in large mmWave MIMO systems reduces to a
sequential detection problem that finds the locations and values
of the non-zero-valued bins in a two-dimensional rectangular grid,
and the optimal adaptive pilot design problem can be cast into the
framework of a partially observable Markov decision process
(POMDP).  Under the POMDP framework, an optimal adaptive pilot
beam sequence
    design method is obtained to maximize the accumulated transmission data rate for a given period of time.
Numerical results are provided to validate
    our pilot signal design method and they show that the proposed method yields good performance.

\end{abstract}


\section{Introduction}

Millimeter-wave (mmWave) communication is rising as a key
technology to provide high data rates with wide bandwidth (BW) in
future wireless systems. However, the signal pathloss in the
mmWave band is much larger than that in the lower band  currently
used in most wireless access networks. To overcome the pathloss,
there is on-going research about highly directional beamforming
techniques in mmWave systems using large antenna
arrays\cite{Ayach&HeathEtAl:12ICC,
Alkhateeb&Ayach&Leus&Heath:13ITA,
Alkhateeb&Ayach&Leus&Heath:14arxiv}. Typically these beamforming
techniques require channel state information (CSI) at the
transmitter and the receiver, but it is more difficult to obtain
CSI in the mmWave band than in the lower band because of the high
propagation directivity and  the low signal-to-noise ratio (SNR)
before beamforming. Thus, the accurate and efficient channel
estimation is important to attain the promised BW gain of the
mmWave band.

One of the major differences between the channels in conventional
MIMO systems in lower bands and large mmWave MIMO systems is the
sparsity in the MIMO channel. Whereas channel estimation methods
in conventional lower-band MIMO systems assume rich scattering or
the knowledge of the channel covariance matrix in the
rank-deficient case, such assumptions are not valid in the mmWave
band \cite{Bajwa&Haupt&Sayeed&Nowak:10IEEE}. Among many possible
ray directions resolved by a large antenna array, only a few
directions actually carry the signal, and these signal-carrying
directions are unknown beforehand
\cite{Bajwa&Haupt&Sayeed&Nowak:10IEEE}. To tackle the challenge of
the sparse channel estimation in the  mmWave band, algorithms
 based on compressed sensing (CS) have recently developed
 \cite{Taubock&Hlawatsch:08ESP, Bajwa&Haupt&Sayeed&Nowak:10IEEE, Alkhateeb&Ayach&Leus&Heath:13ITA, Alkhateeb&Ayach&Leus&Heath:14arxiv}.
In \cite{Bajwa&Haupt&Sayeed&Nowak:10IEEE}, the channel estimation
problem is formulated by capturing the sparse nature of the
channel, and CS techniques are used to analyze the sparse channel
estimation performance. Recently,  an efficient channel estimation
and training beam design method for large mmWave MIMO systems was
proposed based on adaptive CS in
\cite{Alkhateeb&Ayach&Leus&Heath:13ITA,
Alkhateeb&Ayach&Leus&Heath:14arxiv}. In the proposed method, the
channel estimation is conducted over multiple slots under the
assumption that the channel does not vary over the considered
multiple slots. Each slot consists of multiple training beam
symbol times so that the sparse recovery is feasible at each slot,
and the training beam at the next slot is adaptively designed
depending on the previous slot observation based on a space
bisection approach \cite{Alkhateeb&Ayach&Leus&Heath:13ITA,
Alkhateeb&Ayach&Leus&Heath:14arxiv}. Such a design strategy is a
reasonable choice to search the actual signal-carrying directions
in the space.

In this paper, we consider the adaptive pilot beam sequence design
to estimate the sparse channel in large mmWave MIMO systems based
on a decision-theoretical approach, and  propose a strategy to
design the pilot beam sequence that is optimal in a certain sense.
Exploiting the sparsity of mmWave channels with the virtual
channel representation and imposing a Markovian random walk
assumption on the physical movement of the LOS and reflection
clusters, we cast the sparse channel estimation problem in large
mmWave MIMO systems into the framework of a partially observable
Markov decision process (POMDP) with finite horizon
\cite{PutermanBook}, where we need to find the locations and
values of the non-zero-valued bins in a two-dimensional
rectangular grid. Under the proposed POMDP framework, we derive an
optimal adaptive pilot beam sequence design method to maximize the
accumulated transmission data rate for a given period of time.


\section{System Model}
\label{sec:systemmodel}

\subsection{Sparse Channel Modeling in mmWave Systems}

We consider a mmWave MIMO system with a uniform linear array (ULA)
of $N_t$ antennas at the transmitter  and an ULA of $N_r$ antennas
at the receiver.  The received signal at symbol time $n$ is given
by
\begin{equation}
\ybf_n= \Hbf_n \xbf_n + \nbf_n, ~~ n=1,2,\cdots,
\end{equation}
where $\Hbf_n$ is the $N_r\times N_t$ MIMO channel matrix at time $n$, $\xbf_n$ is
the $N_t \times 1$ transmitted symbol vector at time $n$ with a power constraint $\Ebb\{\xbf_n\xbf_n^H\} \le P_t$,
and $\nbf_n$ is the $N_r \times 1$
Gaussian noise vector at time $n$ from  $\Cc\Nc({\bf 0}, \sigma_N^2 \Ibf_{N_r})$.
The MIMO channel matrix $\Hbf_n$ can be expressed in terms of the physical propagation paths as
\begin{equation}\label{eq:physical_channelmodel}
\Hbf_n = \sqrt{N_tN_r}\sum_{\ell=1}^L \alpha_{n,\ell}\abf_{RX}(\theta_{n,\ell}^r)
\abf_{TX}^H(\theta_{n,\ell}^t),
\end{equation}
where $\alpha_{n,\ell} \sim \Cc\Nc(0,\xi^2)$ is the complex gain
of the $\ell$-th path at time $n$, and $\theta_{n,\ell}^r$ and
$\theta_{n,\ell}^t$ are the angle-of-arrival (AoA) and
angle-of-departure (AoD) normalized directions of the $\ell$-th
path at time $n$ for the receiver and the transmitter,
respectively. Here, the normalized direction $\theta$ is related
to the physical angle $\phi \in [-\pi/2,\pi/2]$ as $ \theta =
\frac{d\sin(\phi)}{\lambda}$, where  $d$ is the spacing between
two adjacent antennas and $\lambda$ is the signal wavelength (we
assume $\frac{d}{\lambda} = \frac{1}{2}$), and
$\abf_{RX}(\theta^r)$ and $\abf_{TX}(\theta^t)$ are the receiver
response and the transmitter steering vector, which are defined as
\cite{Sayeed&Raghavan:07JSTSP}
\begin{align}
\abf_{RX}(\theta^r) &= \frac{1}{\sqrt{N_r}}
[1,e^{-\iota 2 \pi \theta^r},\cdots,e^{-\iota(N_r -1) 2 \pi  \theta^r}]^T,\\
\abf_{TX}(\theta^t) &= \frac{1}{\sqrt{N_t}}
[1,e^{-\iota 2 \pi \theta^t},\cdots,e^{-\iota(N_t -1) 2 \pi  \theta^t}]^T.
\end{align}
With neglecting the angle quantization error the physical MIMO channel matrix $\Hbf_n$ can
be rewritten in terms of the virtual channel matrix $\Hbf_{n}^V$ \cite{Sayeed:02SP}:
\begin{equation}
\Hbf_n = \Abf_R \Hbf_{n}^V \Abf_T^H,
\end{equation}
where $\Abf_R =
[\abf_{RX}(\tilde{\theta}^r_{1}),\cdots,\abf_{RX}(\tilde{\theta}^r_{N_r})]$,
$\tilde{\theta}^r_{i}=-\frac{1}{2} + \frac{i-1}{N_r}$ for
$i=1,\cdots,N_r$, and $\Abf_T =
[\abf_{TX}(\tilde{\theta}^t_{1}),\cdots,\abf_{TX}(\tilde{\theta}^t_{N_t})]$,
$\tilde{\theta}^t_{j}=-\frac{1}{2} + \frac{j-1}{N_t}$ for
$j=1,\cdots,N_t$. The element in the $i$-th row and the $j$-th
column of $\Hbf_{n}^V$ indicates the complex channel gain whose
AoA and AoD normalized directions are $\tilde{\theta}_{r,i}$ and
$\tilde{\theta}_{t,j}$, respectively. The sparsity in the physical
channel model in \eqref{eq:physical_channelmodel} is translated
into
\[
\sum_{j=1}^{N_t} ||\Hbf_{n}^V(:,j)||_0 = L,
\]
where $L \ll N_tN_r$ for large $N_t$. We assume that the receiver has $N_r ~(\ll N_t)$ RF chains so that it can implement the filter bank
 $\Abf_R^H$ to look ahead for all possible AoA directions. In this case, the receiver filter-bank output is given by
\begin{equation}\label{eq:filterbank_received}
\ybf_n' := \Abf_R^H\ybf_n= \Hbf_{n}^V\Abf_T^H\xbf_n + \nbf_n'
\end{equation}
where $\nbf_n' = \Abf_R^H\nbf_n$.

By simply transmitting a pilot beam sequence
$\abf_{TX}(\tilde{\theta}^t_{1})$,
$\abf_{TX}(\tilde{\theta}^t_{2})$, $\cdots$,
$\abf_{TX}(\tilde{\theta}^t_{N_t})$, the receiver can estimate the
positions and values of the $L$ non-zero elements of $\Hbf_{n}^V$
if the channel is time-invariant for $N_t$ symbol times. However,
such a method  does not exploit the channel sparsity and/or the
channel dynamic, and is inefficient when $N_t$ is large.

\subsection{The Proposed Dynamic Channel Model}
\label{subsec:FSMCchModel}

To design a very efficient pilot beam sequence, we exploit the
channel dynamic, and model the channel dynamic by using a
Markovian structure that is different from the Gauss-Markov or
state-space channel model conventionally used to model the channel
dynamic. In large mmWave MIMO systems, the sparsity should be
captured in the channel dynamic.  Here, we focus on the locations
of the non-zero elements of $\Hbf_n^V$ rather than the values,
since the value will be obtained with reasonable quality once the
correct direction is hit by the pilot beam with high power.
Note that each propagation path is generated by either LOS or a
reflection cluster and the physical movement of the receiver or a
reflection cluster can be modelled as a random walk in space. This
random walk translates into each nonzero bin's random walk in the
virtual channel matrix. Thus, we assume a stationary block
Markovian random walk for the dynamic of the virtual channel matrix.
That is, the virtual channel matrix $\Hbf_{(k)}^V$  at slot $k$ is
constant over the slot and changes to $\Hbf_{(k+1)}^V$ at the next
slot $k+1$ with the aforementioned random walk with a set of
transition probabilities. We also assume that the movement of each
path is independent. Since the receiver checks all possible AoA
directions in parallel, we here only consider the random walk
across AoD, i.e., the column-wise movement of each non-zero bin in
the virtual channel matrix.

\subsubsection{The Single Path Case}
\label{sec:singlepathcase}

First, consider the single path case, i.e., $L =1$. The single
path (or non-zero bin) is located in a certain column of the
virtual channel matrix at slot $k$, and stay at the same column or
moves to another column of the virtual channel matrix at slot
$k+1$ according to the explained random walk. Since we have $N_t$ columns in the
virtual channel matrix, the number $N$ of states for $L=1$ is
$N_t$. Let us  denote the set of all possible states by
$\mathcal{S}$ given by
\[
\Sc = \{1,2,\cdots, N_t\},
\]
 where  state $i$  denotes
 the state that the path is located in the $i$-th column of the virtual channel matrix.
With the set $\Sc$ of states defined, the $(i,j)$-th element of
the $N \times N$ state transition probability matrix $\Pbf$ is
given by
 \begin{equation}\label{eq:transitionprobability}
    p_{ij} = \text{Pr}\{ S_{k+1} = j | S_k = i\}, ~~ i,j \in
    \mathcal{S},
\end{equation}
where $S_k$ and $S_{k+1}$ denote the states of slots $k$ and
$k+1$, respectively.
 The transition probability matrix captures the
characteristics of the path's movement behavior and thus it should
carefully be designed by considering the physics of the receiver
and reflection cluster movement, e.g., vehicular channels or
pedestrian channels.  Intuitively, it is reasonable to model
$\Pbf$ such that  the transition probability from column $i$ to
column $j$ is monotonically decreasing with respect to (w.r.t.)
$|i-j|$. That is, it is more likely to shift to a nearby column.
In the extreme case of a static channel, we have $\Pbf=\Ibf$.
 If we ignore the possibility of the path's movement
with a large AoD change, we can model the transition probability
matrix as a banded matrix. For the example of $N_t= 5$, one way to
model the transition probability matrix is given by

 {\footnotesize
 \begin{equation}\label{eq:bandedstructure}
    \Pbf_{\beta}^{\text{B}} = \left[\begin{array}{cccccc}
                              1 - \sum_{i=1}^{2} \alpha \beta^i & \alpha \beta & \alpha \beta^2 & 0 & 0 \\
                              1-  \sum_{i=0}^{2} \alpha \beta^i & \alpha & \alpha \beta & \alpha \beta^2 & 0 \\
                              \alpha \beta^2 & \alpha \beta & \alpha & \alpha  \beta & \alpha \beta^2\\
                              0 & \alpha \beta^2 & \alpha \beta & \alpha & 1-  \sum_{i=0}^{2}  \alpha \beta^i\\
                              0 & 0 & \alpha \beta^2 & \alpha \beta & 1 - \sum_{i=1}^{2} \alpha \beta^i\\
                            \end{array}
    \right],
 \end{equation}
 }
 where $\beta$ is a decreasing factor that captures the amount of
 probability reduction w.r.t. the column distance, and $\alpha$ is the value that makes
the sum of each row in \eqref{eq:bandedstructure}  one.

\subsubsection{The Multiple Path Case}
\label{sec:multipathscase}

Now consider the multiple propagation path case, i.e., $L \ge 2$.
In this case, each path (or non-zero bin) is located in a column
of the virtual channel matrix. The set $\Sc$ of all possible states is now given by
\begin{equation}
\Sc=\{(i_1,i_2,\cdots,i_L), ~i_1,i_2,\cdots,i_L =1, 2,\cdots,N_t\},
\end{equation}
where state $(i_1,\cdots,i_\ell,\cdots,i_L)$ denotes that the  $\ell$-th path is located at the $i_\ell$-th column of the virtual channel matrix for $\ell=1,\cdots, L$ and the cardinality $N$ of $\Sc$ is $N_t^L$.  For notational simplicity, let us also use the following notation:
\begin{equation}
\Sc=\{\sbf^{(1)},\sbf^{(2)},\cdots,\sbf^{(N)}\},
\end{equation}
where states $(i_1,\cdots,i_L)$, $i_1,\cdots,i_L=1,\cdots,N_t$, are enumerated into states $\sbf^{(i)}$, $i=1,2,\cdots, N=N_t^L$.
We here allow multiple paths can merge on and diverge from a column of the virtual matrix. Then the state transition probability in the $L$ independent path case is given by
 \begin{align}
   & \text{Pr}\{ S_{k+1} = (j_1,\cdots,j_L) | S_k = (i_1,\cdots,i_L)\} \nonumber \\
   &~~= p_{i_1j_1}p_{i_2j_2} \times \cdots \times p_{i_L j_L}, \label{eq:transitionprobabilityL}
\end{align}
where $p_{ij}$ denotes the transition probability that a path
moves from the $i$-th column to the $j$-th column of the virtual channel matrix at the next slot and is defined in \eqref{eq:transitionprobability}.

\begin{figure}[t]
\begin{psfrags}
        \psfrag{mp5}[c]{\tiny $-0.5$} %
        \psfrag{p5}[c]{\tiny $0.5$} %
        \psfrag{al1}[c]{\tiny $\alpha_{1,1}$} %
        \psfrag{al2}[c]{\tiny $\alpha_{1,2}$} %
        \psfrag{al3}[c]{\tiny $\alpha_{1,3}$} %
        \psfrag{aoa}[c]{\tiny $\{\tilde{\theta}^r_{m}\}$ (AoA)} %
        \psfrag{aod}[c]{\tiny $\{\tilde{\theta}^t_{n}\}$ (AoD)} %
        \psfrag{alk1}[c]{\tiny $\alpha_{2,1}$} %
        \psfrag{alk2}[c]{\tiny $\alpha_{2,2}$} %
        \psfrag{alk3}[c]{\tiny $\alpha_{2,3}$} %
        \psfrag{t1}[c]{\small at time $k=1$} %
        \psfrag{t2}[c]{\small at time $k=2$} %
    \centerline{ \scalefig{0.5} \epsfbox{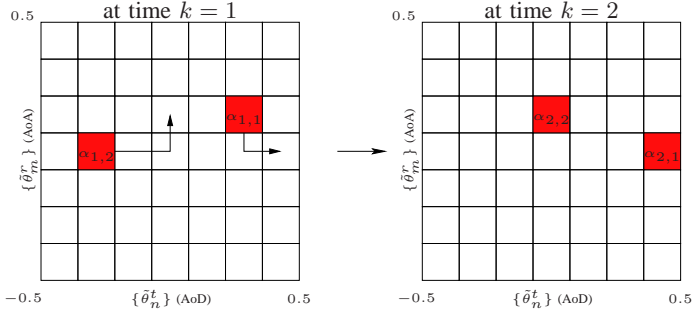} }
    \caption{An illustration of a transition of each path when $L=2$
    and $N_t=N_r=7$. }
    \label{fig:aod_bins}
\end{psfrags}
\end{figure}

Fig. \ref{fig:aod_bins} illustrates an example of transitions of the paths when $L=2$ and $N_t = N_r = 7$.
The transition probability of Fig. \ref{fig:aod_bins} is $p_{67} \times p_{24}$
due to the independence assumption for each path.

\subsection{Channel Sensing with Pilot Beams}

\begin{figure}[t]
\begin{psfrags}
        \psfrag{mp}[c]{\small $M_p $} %
        \psfrag{st}[c]{\small State transition}
        \psfrag{k}[c]{\small  slot $(k)$}
        \psfrag{k1}[c]{\small slot $(k+1)$}
        \psfrag{cd}[c]{\small $\cdots$}
        \psfrag{cs}[c]{\small Channel sensing}
        \psfrag{ce}[c]{\small and estimation}
        \psfrag{rx}[c]{\small RX}
        \psfrag{tx}[c]{\small TX}
        \psfrag{hb}[c]{\small $\Hbf_k$}
        \psfrag{fb}[c]{\small Feedback}
        \psfrag{dp}[c]{\small Data transmission}
        \psfrag{td}[c]{\small $\quad\quad $ A sequence of pilot beams
        for $M_p$ symbol times}
        \psfrag{a1}[c]{\small $\abf_{TX}(\tilde{\theta}^t_{i_1})$}
        \psfrag{a2}[c]{\small $\abf_{TX}(\tilde{\theta}^t_{i_2})$}
        \psfrag{a4}[c]{\small $\abf_{TX}(\tilde{\theta}^t_{i_{M_p}})$}
    \centerline{ \scalefig{0.44} \epsfbox{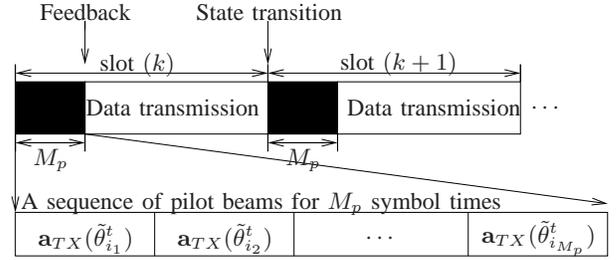} }
    \caption{The process of the pilot training.}
    \label{fig:pilots_model}
\end{psfrags}
\end{figure}

We assume that $M_p$ $(L \le M_p\ll N_t)$ symbol times in each
slot are used for transmitting a sequence of pilot beams and one
column of $\Abf_T$ is selected as the pilot beam in each pilot
symbol time. (We assume that highly directional pilot beam is
required to obtain a channel gain estimate with reasonable quality
due to large pathloss in the mmWave band.) Hence, $M_p$ columns of
$\Abf_T$ are selected as the pilot beam sequence for the $M_p$ pilot
symbol times per slot. If $\abf_{TX}(\tilde{\theta}^t_{i_m})$ is
transmitted as the pilot signal at the $m$-th pilot symbol time in
the $k$-th slot, from \eqref{eq:filterbank_received}, the receiver
filter-bank output is given by
\begin{equation}\label{eq:receivedsig}
\ybf_{(k)}'[m] = \Hbf_{(k)}^V(:,i_m) + \nbf_{(k)}'[m],
\end{equation}
where $\ybf_{(k)}'[m]$ denotes the receiver filter-bank output at
symbol time $m$ of slot $k$,  $\nbf_{(k)}'[m]$ is similarly
defined, and $\Hbf_{(k)}^V(:,i_m)$ denotes the $i_m$-th column of
$\Hbf_{(k)}^V$. After the transmission of the sequence of pilot
beams for one slot is finished, the receiver senses and estimates
the $M_p$ columns of $\Hbf_{(k)}^V$ corresponding to the $M_p$
pilot beams. Then, the receiver feedbacks the sensing results and
estimated channel gains corresponding to the $M_p$ pilot beam
directions to the transmitter. The process is depicted in Fig.
\ref{fig:pilots_model}.

Now, the problem is to design the sequence of pilot beams for
$M_p$ symbol times for each slot in a certain optimal way.
 Since $M_p \ll N_t$, we can only sense a few columns of  $\Hbf_{(k)}^V$ at slot $k$.
  Therefore, $M_p$ pilot beams at each slot should be designed
  judiciously by exploiting the channel dynamic and the available
  information in all the previous slots.

\section{POMDP Formulation for Pilot Beam Design for Sparse Channel Estimation}
\label{sec:POMDPformulation}

\subsection{Action Space at the Transmitter and Feedback from the Receiver}
\label{sec:actionandobservation}

In Section \ref{sec:systemmodel}, we assumed that $M_p$ columns of $\Abf_T$ are selected as the pilot beam sequence
for the $M_p$ pilot symbols per slot. This is equivalent to choose $M_p$ columns of the virtual channel matrix at each slot
to be sensed by the pilot beam sequence. We denote the selected column indices of the virtual channel matrix
by
\[
\abf = [a_1, a_2, \cdots, a_{M_p}],
\]
 where $a_m$ indicates the index of the column of the virtual channel matrix
that is sensed at symbol time $m$. ($\abf$ is referred to as the action vector.) Hence, there are ${N_t}\choose{M_p}$ possible $\abf$'s
and the optimal pilot beam sequence design problem reduces to choosing the
best $\abf$ at each slot.

After the chosen pilot beam sequence is transmitted to the receiver,
the receiver feedbacks the result of detection  to the transmitter for
pilot beam sequence design for the next slot.
The feedback information contains the information about the existence\footnote{The detection can be wrong. This is another reason for POMDP in addition to the limited search of $M_p$ columns out of the $N_t$ total columns per slot.} of
paths in the selected columns of the virtual channel matrix  as well as the complex gains of the
detected paths. Then, the transmitter uses the channel gain information of the detected paths for beamforming during the data transmission period and uses
the feedback information about the existence of paths to choose the pilot beam sequence for the next slot in an adaptive manner.
The latter feedback information can be modeled as
\begin{equation}
    \obf = [o_1, o_2, \cdots, o_{M_p}] \in \{0,1\}^{M_p},
\end{equation}
where $o_m = 1$ indicates that a path is detected by
the pilot beam transmitted at the $m$-th pilot symbol time, and otherwise $o_m =0$.
Since there exist $2^{M_p}$ possibilities in $\obf$, the feedback information
space is  defined as ${\mathcal{O}}= [\obf^{(1)}, \obf^{(2)}, \cdots, \obf^{(2^{M_p})}]$.
When the current state of the virtual channel matrix is $\sbf^{(i)}$ and the action vector $\abf$ is selected for
the pilot beam sequence for the current slot, the probability that the transmitter
observes the feedback information  $\obf^{(j)}$ is denoted as $q_{ij}^\abf$, i.e.,
\begin{align}
     &q_{ij}^{\abf} \triangleq \text{Pr}\{ \obf = \obf^{(j)} | \sbf^{(i)}, \abf\} ~~~\text{for}~ \sbf^{(i)} \in {\mathcal{S}}, \obf^{(j)} \in \mathcal{O}. \label{eq:Pr_feedbackInf}
\end{align}
This probability depends on the detector used to identify the existence of a non-zero bin in a column at the receiver.

\subsection{Sufficient Statistic}
\label{sec:sufficientstatistic}

  \begin{figure}[t]
\begin{psfrags}
        \psfrag{block1}[c]{\small slot $k$} %
        \psfrag{a1}[c]{\footnotesize $\Pbf$}
        \psfrag{a2}[c]{\footnotesize $\abf$}
        \psfrag{a3}[c]{\footnotesize $\obf$}
        \psfrag{a4}[c]{\footnotesize $r_k(\sbf,\abf,\obf)$}
        \psfrag{bleif}[c]{\footnotesize $\pibf_k$}
        \psfrag{next}[c]{\footnotesize $\pibf_{k+1}$}
        \psfrag{b1}[c]{\footnotesize $\text{State}$}
        \psfrag{b2}[c]{\footnotesize $\text{transition}$}
        \psfrag{c1}[c]{\footnotesize $\text{Pilot beam}$}
        \psfrag{c2}[c]{\footnotesize $\text{selection}$}
        \psfrag{d1}[c]{\footnotesize $\text{Feedback}$}
        \psfrag{d2}[c]{\footnotesize $\text{observation}$}
        \psfrag{e1}[c]{\footnotesize $\text{}$}
        \psfrag{e2}[c]{\footnotesize $\text{Reward}$}
    \centerline{ \scalefig{0.38} \epsfbox{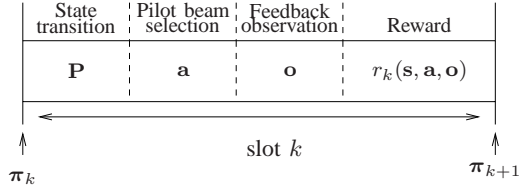} }
    \caption{The sequence of operation at slot $k$}
    \label{fig:Thesequenceofoperation}
\end{psfrags}
\end{figure}

Fig. \ref{fig:Thesequenceofoperation} describes the sequence of the operation
at slot $k$ consisting of the current belief vector $\pibf_k$, state transition, action, observation, and reward.
At the beginning of slot $k$,  the information from all the past slot pilot beam sequences and feedback information
can be summarized as a belief vector:\footnote{Note that the belief vector $\pibf_k$ is conventionally defined prior to the state transition for each slot, as shown in Fig. \ref{fig:Thesequenceofoperation}. The belief vector after the state transition can simply be updated by using the state transition probability matrix.}
\begin{equation}
    \pibf_k = [\pi_{k,1}, \pi_{k,2}, \cdots, \pi_{k,N}],
\end{equation}
where  $\pi_{k,i}$
is the probability that the state at the beginning of slot $k$ is state $\sbf^{(i)}$ conditioned on all past pilot beam sequences and feedback
information.  It is known
that the belief vector is a sufficient statistic for the action, i.e., the design of the optimal
pilot beam sequence for slot $k$ \cite{Smallwood&Sondic:73OR}.
The transmitter uses the belief vector
to optimally choose the action, i.e., the pilot beam sequence for slot $k$ that maximizes the expected reward,
and updates the belief vector for the next block based on new feedback information.

\subsection{The Reward and The Policy}
\label{sec:therewardandpolicy}

A reward is gained during the data transmission period according to the accuracy of channel estimation.
According to the objective, there can be several
ways to define the reward. Since we want to track all actual propagation paths in the sparse mmWave MIMO channel successfully,
we define the reward for each slot as the number\footnote{With $\alpha_{(k),l} \stackrel{i.i.d.}{\sim} \Cc\Nc(0,\xi^2)$ in \eqref{eq:physical_channelmodel}, the data rate will roughly be $\log (1 + N_p \xi^2)$ in case of maximal ratio combining (MRC) transmission  or $N_p \log (1 + \xi^2)$ in case of spatial multiplexing used at the transmitter, where $N_p$ is the number of identified paths. Thus, $N_p$ directly captures the data rate for the case of spatial multiplexing. In case of MRC transmission, $\log (1+N_p \xi^2)$ can be used as the reward without change in the formulation.}  of actual propagation paths (i.e., the number of non-zero bins in the virtual channel matrix) detected by
the selected pilot beam sequence. If the state of the virtual channel matrix at slot $k$
is $\sbf^{(i)}$,  the selected pilot beam sequence or the action vector at slot $k$ is $\abf$, and the transmitter observes
the feedback information $\obf^{(j)}$, then the immediate reward at slot $k$ is expressed as
\begin{equation}
    r(\sbf^{(i)},\abf,\obf^{(j)}) = \sum_{m=1}^{M_p} N^{BIN}_{\sbf^{(i)}, a_m}  o_{m}^{(j)}
\end{equation}
where $N^{BIN}_{\sbf^{(i)}, a_m}$ is the number of non-zero bins in column $a_m$ when the virtual channel matrix is in state $\sbf^{(i)}$, and $o_m^{(j)}$ is the $m$-th element of $\obf^{(j)}$. (Two paths with the same AoD and different AoAs are considered as two different paths.)   Here, the false alarm of the receiver detector does not affect the immediate reward because in this case $o_{m}^{(j)}=1$ but $N^{BIN}_{\sbf^{(i)}, a_m}=0$. In the case of miss detection, the opportunity is simply lost.

Since the state at slot $k$ and the feedback information
are unknown at the time of action, we should consider the expected reward \cite{PutermanBook}.
If the state of the virtual channel matrix prior to the
state transition at slot $k$ is $\sbf^{(n)}$, then the immediate expected
reward at slot $k$ is
\begin{eqnarray}
  \nonumber &&  R(\sbf^{(n)},\abf) \\
  \nonumber &&  = \sum_{i=1}^N p_{ni}\sum_{j=1}^{2^{M_p}}\text{Pr}\{ \obf = \obf^{(j)} |  \sbf^{(i)}, \abf\}  r(\sbf^{(i)},\abf,\obf^{(j)}) \\
            &&  = \sum_{i=1}^N p_{ni}\sum_{j=1}^{2^{M_p}} q_{ij}^\abf \sum_{m=1}^{M_p}N^{BIN}_{\sbf^{(i)}, a_m} o_{m}^{(j)},
\end{eqnarray}
where the state transition from $\sbf^{(n)}$ to all possible $\sbf^{(i)}$ within the slot is captured by $\sum_{i=1}^N p_{ni} (\cdot)$.
When the belief vector $\pibf_k$ is given at the beginning of slot $k$ and $S_k$ is the random variable representing the state at the beginning of the slot, the immediate expected reward at slot $k$ is given by
\begin{eqnarray}
  \nonumber  {\mathcal{R}}(\pibf_k , \abf ) &=& {\mathbb{E}}\{R(S_k, \abf) | \pibf_k\} \\
                          &=& \langle\Rbf(\abf),  \pibf_k\rangle
\end{eqnarray}
where  $\Rbf(\abf) = [ R(\sbf^{(1)},\abf),  R(\sbf^{(2)},\abf), \cdots,  R(\sbf^{(N)},\abf)]$,
and $\langle \cdot,\cdot \rangle$ denotes the inner product operation.

In the POMDP framework, a policy $\delta$ is defined as a sequence of functions that
 maps the belief vector to an action for each slot \cite{Smallwood&Sondic:73OR,Monahan:82MS}, where the action in our formulation is the choice of
the pilot beam sequence. The optimal policy is the policy that
maximizes the total immediate expected reward over $T$ slots\footnote{Such a formulation is called a finite-horizon POMDP. The formulation here can be modified to the infinite-horizon case.} when
the initial belief vector $\pibf_1$ at the beginning of the transmission is given. In other words, the optimal policy  $\delta^*$ is expressed as
\begin{equation}\label{eq:optimalreward}
    \delta^{*} = \mathop{\arg\max}_{\delta} {\mathbb{E}}_{\delta}\left[\sum_{k=1}^T R(S_k,\abf_k)|\pibf_1\right],
\end{equation}
where $\abf_k$ is the action vector at slot $k$, ${\mathbb{E}}_\delta$ is the conditional expectation when the policy $\delta$ is given,
and $T$ is the total number of slots.

\section{The Optimal and Suboptimal Strategies for Channel Estimation}
\label{sec:strategy}

Under the proposed formulation the optimal pilot beam sequence design problem is equivalent to the problem of finding
the optimal policy that satisfies \eqref{eq:optimalreward}.
When the initial belief vector $\pibf_1$ is given at $k=1$, we can define an {\it
optimal  value function} $V(\pibf_1)$ as the maximum total expected reward
by the optimal policy $\delta^*$ :
\begin{equation}
    V(\pibf_1) = {\mathbb{E}}_{\delta^*}\left[ \sum_{k=1}^T R(S_k,\abf_k)|\pibf_1 \right].
\end{equation}
Now consider $V^{k}(\pibf_k)$ defined as the maximum remaining expected reward that can be obtained from
slot $k$ to slot $T$ when a belief vector $\pibf_k$ is given at the beginning of slot $k$.
Then, $V^{k}(\pibf_k)$  can be decomposed as  \cite{Ross70applied}
{\footnotesize
\begin{align}\label{eq:optimalvaluefunction}
    \nonumber & V^k(\pibf_k)  \\
              & = \max_{\abf_k} \left\{ \langle \Rbf(\abf_k), \pibf_k \rangle  + \sum_{j=1}^{2^{M_p}} V^{k+1}({\mathcal{T}}(\pibf_k | \obf^{(j)}, \abf_k)) \gamma(\obf^{(j)}| \pibf_k, \abf_k)  \right\},
\end{align}
}

\noindent where $\gamma(\obf^{(j)}| \pibf_k, \abf_k) = \sum_{i=1}^N q_{ij}^{\abf_k} \sum_{n=1}^N \pi_{k,n} p_{ni}$
and ${\mathcal{T}}(\pibf_k | \obf^{(j)}, \abf_k)$ is the updated belief vector from $\pibf_k$ at
slot $k$ for the next slot after taking action $\abf_k$ and observing $\obf^{(j)}$. ${\mathcal{T}}(\pibf_k | \obf^{(j)}, \abf_k)$ can easily be computed using Bayes's formula as \cite{Smallwood&Sondic:73OR,Monahan:82MS}
\begin{equation}\label{eq:beliefupdate}
   \pibf_{k+1} = [\pi_{k+1, 1}, \pi_{k+1, 2}, \cdots, \pi_{k+1, N}] =  \nonumber {\mathcal{T}}(\pibf_k | \obf^{(j)},\abf_k),
\end{equation}
where the $i$-th element of $\pibf_{k+1}$ is given by
\begin{align}\label{eq:beliefupdate1}
 \pi_{k+1,i} &= \text{Pr}\{ \sbf_k = \sbf^{(i)} | \obf^{(j)},\abf_k,\pibf_k\} \nonumber \\
                                               &= \frac{q_{ij}^{\abf_k} \sum_{n=1}^N \pi_{k,n} p_{ni}}{\sum_{i=1}^N q_{ij}^{\abf_k} \sum_{n=1}^N \pi_{k,n}p_{ni}}.
\end{align}

The optimal policy $\delta^*$ over the considered transmission period $k=[1,2,\cdots,T]$ can be computed via the recursion \eqref{eq:optimalvaluefunction} from backward once the state transition probability $\Pbf$, reward, and observation and action spaces are given  \cite{Smallwood&Sondic:73OR,Monahan:82MS,Cassandra&Anthony:97UAI}. This computation can be done off-line and the optimal policy can be stored beforehand.\footnote{Hence, for each channel type, we can pre-compute the policy and store it. This is one of the main advantages of the proposed pilot beam design approach.}  Then, in actual transmission, we start from $k=1$ with $\pibf_1$ and repeat action and observation. The complexity of this actual operation is insignificant.

As the number of states and the size of the action space increase, obtaining the optimal policy for the POMDP problem requires high  complexity. To reduce the computational complexity, we can use point-based POMDP value iteration algorithms
proposed in \cite{Pineau&Gordon:03IJCAI,Kurniawati&Hsu:08Robtics,Smith&Simmons:12arXiv}.
Alternatively, to reduce the complexity, we can simply use the greedy policy that considers only the immediate expected reward at
each slot.

\section{Numerical results}
\label{sec:NumericalResult}

In this section, we provide some numerical results to evaluate  the performance
of the proposed pilot beam design and channel estimation for sparse mmWave MIMO channels. Throughout the simulation, we fixed the number of paths
in the MIMO channel as $L=2$ and assume that each path moves independently.
The receiver uses the ML detector for sensing the paths and the noise power
at the receiver side is $\sigma_N^2 =1$ and the transmit power is fixed to $1$ dB
higher than the noise power before beamforming. (After beamforming with $N_t$ transmit antennas, the SNR increases correspondingly.)

\begin{figure}[tp]
\centerline{ \SetLabels
\L(0.5*-0.1) (a) \\
\endSetLabels
\leavevmode
\strut\AffixLabels{
\scalefig{0.41}\epsfbox{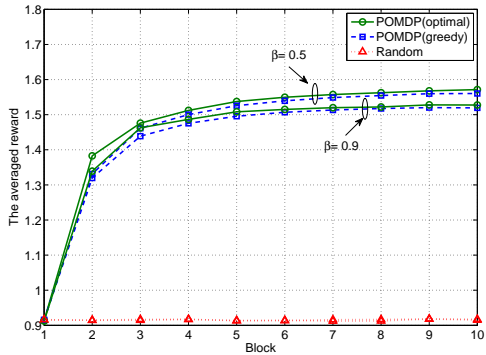}
 } } \vspace{0.5cm}
\centerline{ \SetLabels
\L(0.5*-0.1) (b) \\
\endSetLabels
\leavevmode
\strut\AffixLabels{
\scalefig{0.41}\epsfbox{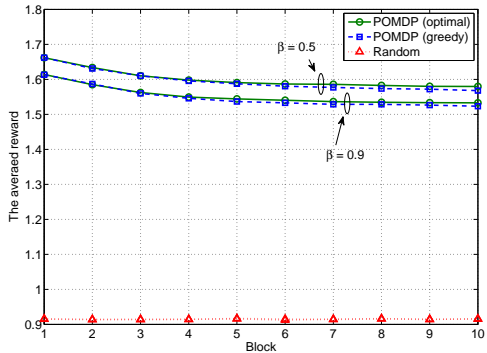}
 } } \vspace{0.5cm}
\caption{The performance of the optimal strategy and the greedy strategy in POMDP. ($N_t =8, N_r = 4, M_p = 4, L =2 $)}
\label{fig:Averagedreward}
\end{figure}

We first considered the MIMO channel with $N_t=8$ transmit antennas and $N_r =4$ receive antennas.
We set $M_p =4$ and used $\Pbf_{\beta}^B$ in \eqref{eq:bandedstructure} with bandwidth $1$ for
each path's movement. Fig. \ref{fig:Averagedreward} shows the performance of the optimal POMDP strategy, the greedy POMDP strategy and
a random beam selection strategy, where the $y$-axis in the figure indicates the slot reward averaged over  $100,000$ MIMO channel process realizations.
  Fig. \ref{fig:Averagedreward} (a) shows the performance when  the initial
state information is unknown so that all the elements of the initial belief vector are set to be equal, and
Fig. \ref{fig:Averagedreward} (b) shows the performance when the initial
state information is perfectly known. Thus, we can consider that Fig. \ref{fig:Averagedreward} (a) shows the initial transient behavior whereas Fig. \ref{fig:Averagedreward} (b) shows the steady-state tracking behavior.
In the random pilot beam selection strategy, one of ${N_t}\choose{M_p}$ possible pilot beam sequences is
randomly selected to sense the channel at each slot.
It is shown in Fig. \ref{fig:Averagedreward} that the optimal POMDP strategy and the greedy PODMP strategy
significantly outperform the random pilot beam selection strategy. It is also shown in the figure that the performance gap between the optimal strategy
and the greedy strategy is not significant in this case.

\begin{figure}[!t]
\centering
\scalefig{0.4}\epsfbox{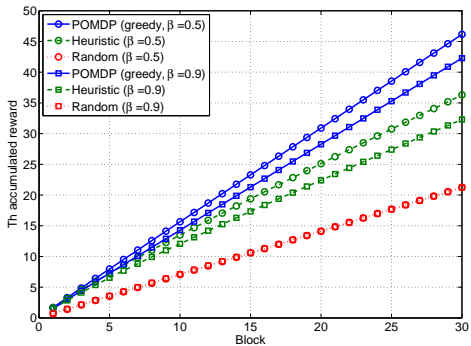}
\caption{The performance of the greedy strategy in POMDP, a heuristic tracking strategy in the MIMO channel,
and the random selection strategy. ($N_t =16, N_r = 4, M_p = 6, L =2$),}
\label{fig:accumulated_reward_middleantennaarray}
\end{figure}

Next, we considered a larger MIMO channel with $N_t=16$ transmit antennas and $N_r=4$ receive antennas.
We set $M_p =6$ and used $\Pbf_{\beta}^B$  in \eqref{eq:bandedstructure}  with bandwidth $2$ for the channel dynamic.
In the case, we focused on the tracking performance with the assumption of perfect knowledge of the initial state information to show the ineffectiveness of simple heuristic tracking strategies.  In this case, the POMDP size is large and thus we considered the suboptimal greedy pilot beam design method only. Fig. \ref{fig:accumulated_reward_middleantennaarray} shows the performance of
 the greedy POMDP strategy, a heuristic strategy, and the random strategy.
The $y$-axis in Fig. \ref{fig:accumulated_reward_middleantennaarray}  is the averaged accumulated reward to the point of the $x$-axis averaged over $10,000$ MIMO channel process realizations.

In the heuristic tracking strategy, $3$ pilot symbol times out of the total $6$ pilot symbol times per slot  are used to track
one path and the other three symbol times are used to track the other path based on the detection result
of the previous slot.
If a path is detected at a certain column at the previous slot, the algorithm computes the most probable three columns for the next slot pilot beams based on the currently detected column and $\Pbf_{\beta}^B$. If no path is detected at the previous slot, the pilot beam indices do not change.  It is seen that the greedy POMDP strategy yields  better performance than the other strategies.

\section{Conclusion}
\label{sec:conclusion}

We have considered the pilot beam sequence design
for sparse large mmWave MIMO channels.
We have shown that the pilot beam design problem can be formulated as a POMDP problem by exploiting the sparse channel dynamic
and have obtained the optimal strategy and a greedy strategy for pilot beam sequence design. The proposed pilot design method can be used to estimate the channel initially at the beginning of the transmission or to track the channel once the sparse channel locations are identified with some other method. For the initial channel identification purpose, the proposed algorithm can be modified by considering a superposed pilot beam for a pilot symbol time with adaptive resolution over slots to shorten the time for initial path location identification.

\end{document}

%% file: macros.tex
\def\psfancypar#1#2{\begingroup\def\par{\endgraf\endgroup\lineskiplimit=0pt}
               \setbox2=\hbox{\large\sc #2}
               \newdimen\tmpht \tmpht \ht2 \advance\tmpht by \baselineskip
               \font\hhuge=Times-Bold at \tmpht
               \setbox1=\hbox{{\hhuge #1}}
               \count7=\tmpht \count8=\ht1
               \divide\count8 by 1000 \divide\count7 by \count8 
               \tmpht=.001\tmpht\multiply\tmpht by \count7 
               \font\hhuge=Times-Bold at \tmpht
               \setbox1=\hbox{{\hhuge #1}}
               \noindent
                \hangindent1.05\wd1
               \hangafter=-2 {\hskip-\hangindent
               \lower1\ht1\hbox{\raise1.0\ht2\copy1}%
                \kern-0\wd1}\copy2\lineskiplimit=-1000pt}

\newcommand{\E}{\mbox{{\rm E}}}

\newcommand{\abf}{\mbox{${\bf a}$}}

\def\boxit#1{\vbox{\hrule\hbox{\vrule\kern3pt
        \vbox{\kern3pt#1\kern3pt}\kern3pt\vrule}\hrule}}

\def\reals{ { {\rm  I \kern-0.15em R }  } }
\def\complex{ {\,{{\rm C} \kern-0.50em \raise0.20ex {  |}}\, }}

\def\pibf{\hbox{\boldmath$\pi$\unboldmath}}

\def\abf{{\bf a}}

\def\nbf{{\bf n}}
\def\obf{{\bf o}}

\def\sbf{{\bf s}}

\def\xbf{{\bf x}}
\def\ybf{{\bf y}}

\def\xbf{{\bf x}}
\def\ybf{{\bf y}}
\def\Abf{{\bf A}}

\def\Hbf{{\bf H}}
\def\Ibf{{\bf I}}

\def\Pbf{{\bf P}}

\def\Rbf{{\bf R}}

\def\Cc{{\cal C}}

\def\Nc{{\cal N}}

\def\Sc{{\cal S}}

\def\be{\vskip .3cm \begin{equation}}
\def\ee{\end{equation} \vskip .4cm \noindent}
\newcommand{\R}{\mbox{$\hat {\bf R}_{N}$}}

\def\Rxx{\Rbf_{\ssstyle X\kern-.1em X}}

\let\ssstyle=\scriptscriptstyle

\def\Kout{\setbox1=\hbox{\Huge\bf K}\hbox to
1.05\wd1{\hspace{.05\wd1}
}}
\def\Sout{\setbox1=\hbox{\Huge\bf S}\hbox to 1.05\wd1{\hspace{.05\wd1}
}}

%% file: labelfig.tex
  \ifx\LabelFigloaded\MYundefined\relax
  \else
   \fi

  \def\LabelFigloaded{\relax}

  \chardef\LabelFigCatAt\the\catcode`\@
  \catcode`\@=11

 \let\LabelFigwlog@ld\wlog
 \def\wlog#1{\relax}

 \ifx\\\MYundefined@
    \let\\\relax
 \fi

  \def\ms@g{\immediate\write16}

 \def\N@wif{\csname newif\endcsname }
 \def\Temp@ {\N@wif\ifIN@}
 \ifx\INN@\MYundefined@
    \else \let\Temp@\relax
 \fi
 \Temp@

  \def\IN@{\expandafter\INN@\expandafter}
  \long\def\INN@0#1@#2@{\long\def\NI@##1#1##2##3\ENDNI@
    {\ifx\m@rker##2\IN@false\else\IN@true\fi}%
     \expandafter\NI@#2@@#1\m@rker\ENDNI@}
  \def\m@rker{\m@@rker}
 
  \newtoks\Initialtoks@  \newtoks\Terminaltoks@
  \def\SPLIT@{\expandafter\SPLITT@\expandafter}
  \def\SPLITT@0#1@#2@{\def\TTILPS@##1#1##2@{%
     \Initialtoks@{##1}\Terminaltoks@{##2}}\expandafter\TTILPS@#2@}

 \def\Shifted@@#1#2#3{\setbox0=\hbox{#3}%
   \raise -\dp0\vbox {\kern-#2%
       \hbox {\kern#1\unhbox0\kern-#1}%
           \kern#2}}

 \newcount\gridcount
 \newbox\auxGridbox@ \newbox\hGridbox@ \newbox\vGridbox@
 \newbox\Labelbox@ \newbox\auxLabelbox@
 \newbox\Coordinatebox@
 \newtoks\Labeltoks@
 \newdimen\Wdd@ \newdimen\Htt@
 \newdimen\Wddd@ \newdimen\Httt@
 
 \def\Wr@{\immediate\write16}

 \newdimen\GL@wd
 \def\GridLineWidth#1{\GL@wd=#1}

 \def\gobble#1{}
 \def\EdgeErr@{\Wr@{}%
      \Wr@{\string\Edges\space argument
      1, 10, 100 or 1000 please\string!}%
      }

 \newcount\Edgect@

 \def\Sweepup#1\endSweepup{}

 \def\SetEdges@{%
    \edef\Zr@@s{\expandafter\gobble\number\Edgect@\empty}%
        \count255=0\Zr@@s\relax
        \ifnum\count255=\z@\else\EdgeErr@\show\tailtest\fi
        \count255=1\Zr@@s\relax
        \ifnum\count255=\Edgect@\relax\else\EdgeErr@\show\leadtest\fi
    \EdgGl@b\edef\Zr@s{\expandafter\gobble\Zr@@s\empty}
    \ifnum\Edgect@>\@ne\relax\EdgGl@b\let\L@Dc\empty
        \else\EdgGl@b\edef\L@Dc{\string.}\fi
    \ifnum\Edgect@>\@ne\relax
        \EdgGl@b\edef\Edgescale@##1{\divide##1 by \Edgect@}%
        \else\EdgGl@b\edef\Edgescale@##1{}\fi
    }

 \def\Edges#1{\Edgect@=#1\relax
     \let\EdgGl@b\global \SetEdges@}

 \Edges{1}

 \def\hhrule{\hrule height \GL@wd\vskip-.\GL@wd}

 \def\hRule@{%
   \advance\gridcount -2%
   \vfil\hhrule\vfil
   \llap{\smash{\raise -2.5pt
     \hbox{\L@Dc\number\gridcount\Zr@s\kern2pt}}}%
   \hhrule
   }

\def\vvrule{\vrule width \GL@wd \kern-\GL@wd}

 \def\vRule@{\advance\gridcount 2%
   \hfil\vvrule\hfil
   \setbox\auxGridbox@=\vbox to 0pt
      {\vskip \Htt@\vskip 2pt
        \hbox to 0pt{\hss\L@Dc\number\gridcount\Zr@s\hss}\vss}%
      \wd\auxGridbox@=0pt \box\auxGridbox@
   \vvrule
   }

 \def\PlaceGrid@@{\gridcount=10 
  \setbox\hGridbox@=\hbox{%
        \hbox{%
             \hskip-.4pt\vrule
             \vbox to \Htt@{%
               \offinterlineskip\parindent=\z@\relax
               \hbox to \Wdd@{\hfil}
               \hRule@\hRule@\hRule@\hRule@
               \vfil\hhrule\vfil}%
             \vrule\hskip-.4pt}
    }%
  \gridcount=0%
  \setbox\vGridbox@=\hbox{%
      \vbox{\offinterlineskip\parindent=0pt\hsize=0pt
         \vskip-.4pt\hrule%
         \hbox to \Wdd@{%
                 \vtop to \Htt@{\vfil}%
                 \vRule@\vRule@\vRule@\vRule@
                 \hfil\vvrule\hfil}%
         \hrule\vskip-.4pt}}%
  \wd\hGridbox@=0pt\ht\hGridbox@=0pt
  \wd\vGridbox@=0pt\ht\vGridbox@=0pt
  \hbox{\box\hGridbox@\box\vGridbox@}%
  }

 \def\LabelsGlobal{\def\LabGl@b{\global}}
 \def\LabelsLocal{\def\LabGl@b{}}
 \LabelsGlobal 

 \def\SetLabels#1\endSetLabels{%
   \LabGl@b\Labeltoks@={#1()\\}%
   }

 \LabGl@b\Labeltoks@={()\\}

 \def\ShowGrid{\LabGl@b\let\PlaceGrid@\PlaceGrid@@}
 \def\HideGrid{\LabGl@b\let\PlaceGrid@\relax}
 \def\Grids{\ShowGrid\LabGl@b\let\GridSwitch@\ShowGrid}
 \def\noGrids{\HideGrid\LabGl@b\let\GridSwitch@\HideGrid}

 \noGrids

 \def\bAdjust@@{%
     \setbox\auxLabelbox@=\hbox{\raise \dp\auxLabelbox@
            \box\auxLabelbox@}}
 \def\bAdjust@{\let\vAdjust@\bAdjust@@}

 \def\eAdjust@@{\dimen0=-.5\ht\auxLabelbox@
     \advance\dimen0 by .5\dp\auxLabelbox@
     \setbox\auxLabelbox@=
            \hbox{\raise\dimen0\box\auxLabelbox@}}
 \def\eAdjust@{\let\vAdjust@\eAdjust@@}

 \def\tAdjust@@{%
     \setbox\auxLabelbox@=\hbox{\raise-\ht\auxLabelbox@
            \box\auxLabelbox@}}
 \def\tAdjust@{\let\vAdjust@\tAdjust@@}

 \let\vAdjust@\relax

 \def\lAdjust@{\let\hAdjust@\rlap}
 \def\rAdjust@{\let\hAdjust@\llap}

 \let\hAdjust@\relax\let\vAdjust@\relax

 \def\FetchLabel@#1(#2)#3\\{%
     \IN@0#2@@\ifIN@
        \setbox0=\hbox{\ignorespaces#1#3\unskip}%
        \ifdim\wd0>0pt
           \ms@g{}%
           \ms@g{ !!! Bad label(s)? !!!}%
           \message{ #1(#2)#3}%
        \fi
        \def\LabelMole@##1\endFetchLabel@{%
            \IN@0()\\@##1@%
            \ifIN@\def\Temp@{\FetchLabel@##1\endFetchLabel@}%
            \else\def\Temp@{}%
            \fi
            \Temp@
           }%
     \else
       \ignorespaces#1\unskip
       \setbox\auxLabelbox@=%
         \hbox to 0pt{\hss\ignorespaces\hAdjust@
          {\ignorespaces#3\unskip}\hss}%
       \vAdjust@
       \let\hAdjust@\relax\let\vAdjust@\relax
       \AugmentLabelBox@@{#2}%
       \ht\Labelbox@=0pt\dp\Labelbox@=0pt
       \let\LabelMole@\FetchLabel@%
     \fi\LabelMole@}

 \newtoks\XYSep@ 
 \def\SetXYSeparator#1{%
     \IN@0#1@@\ifIN@\XYSep@{*}%
     \else
     \XYSep@{#1}%
     \fi
     }

 \SetXYSeparator*

 \def\AugmentLabelBox@@#1{%
     \IN@0\the\XYSep@ @#1@\ifIN@
       \SPLIT@0\the\XYSep@ @#1@%
       \setbox\Labelbox@=\hbox to 0pt{%
         \unhbox\Labelbox@
         \Shifted@@{\the\Initialtoks@\Wddd@}%
         {\the\Terminaltoks@\Httt@}%
         {\box\auxLabelbox@}}%
     \else
         \ms@g{}%
         \ms@g{ !!! Bad insertion point. !!!}%
         \message{ (#1\ this point was rejected.)}%
     \fi
    }

 \def\FetchOption@#1[#2]#3\endFetchOption@{%
    \def\temp{#1}
    \ifx\temp\empty
       \Edgect@=#2\relax
       \let\EdgGl@b\relax
       \SetEdges@
       \Cleaner@#3%
    \fi}

 \def\Cleaner@#1[@]{\Labeltoks@{#1}}
     
 \def\PlaceLabels@@{\mathsurround=0pt
     \def\Cr@{\\}%
     \let\L\lAdjust@\let\R\rAdjust@
     \let\B\bAdjust@\let\E\eAdjust@\let\T\tAdjust@
     \expandafter\FetchOption@\the\Labeltoks@[@]\endFetchOption@
     \Wddd@=\Wdd@ \Edgescale@\Wddd@ 
     \Httt@=\Htt@ \Edgescale@\Httt@
     \expandafter\FetchLabel@\the\Labeltoks@\endFetchLabel@
     \box\Labelbox@
     }%

 \let \PlaceLabels@\PlaceLabels@@

 \def\AffixLabels#1{\setbox\Coordinatebox@=\hbox{#1}%
      \Wdd@=\wd\Coordinatebox@ \Htt@=\ht\Coordinatebox@
      \advance\Htt@ \dp\Coordinatebox@
      \hbox{\copy\Coordinatebox@\kern-\Wdd@ 
           \Shifted@@{0pt}{-\dp\Coordinatebox@}%
           {\PlaceLabels@\PlaceGrid@}%
           \kern\Wdd@}%
      \GridSwitch@ 
      \LabGl@b\Labeltoks@{()\\}%
      }
 
   \let\wlog\LabelFigwlog@ld   
   \catcode`\@=\LabelFigCatAt  

                                By

              Raymond S\'eroul <A18645@FRCCSC21.BITNET>
                                and 
              Laurent Siebenmann <lcs@topo.math.u-psud.fr>
    
              VERSIONS: July 1991, Oct 1991, Jan 1992, July 1992

INTRODUCTION

      This labelling package is intended for TeX users who
rely on non-TeX sources for for their graphics inserts.  It
provides means for adding TeX labels to such inserts with a
minimum of fuss. 

       For most labels, TeX users have in the past found it
reasonably convenient to rely on non-TeX sources. Typical
occasions when an inescapable need for TeX labels seemed to
arise are

 (a) when the graphics program lacks certain exotic or complex
mathematical symbols

 (b) when the very highest typographical quality is wanted for the
labels

 (c) when labels included with the graphics fail to print, 
 and you cannot figure out why (cf. boxedeps.doc).  The labels
 provided by labelfig.tex are 100

       Since this package first appeared, many users, who in the
past scarcely dreamed of using TeX labels, have come to use
nothing but.  So it is now appropriate to add

Intoxication Warning:  TeX labels may be addictive and expensive. 

     If you have a fast preview you may disagree, and even find
that this package provides an agreeable paste-up environment; see
extra applications at end.

     Note to publishers: It is possible and convenient to ultimately
export the TeX labels produced by labelfig.tex to become an integral
part of the EPS file. This is often desired by a publisher who typically
uses an "upmarket" graphics or page layout program, with which the
staff is skilled in perfecting figures.  See Appendix I for
a recipe.

     The authors are grateful to Patrick Ion of Math Reviews for
helpful comments and encouragement.

BASIC INSTRUCTIONS

    After reading in the macro file using

\input labelfig.tex
preview or proof your figure with a coordinate grid printed on
top, by typing the following:

    \ShowGrid  
    \AffixLabels{<the graphics insertion>}

Here <the graphics insertion> is what you would type to insert
the graphics object alone without the grid.  This must provide
for the space around it. For example <the graphics insertion>
might well be \BoxedEPSF{MyFigure scaled 700} using the
boxedeps.tex macro package (from same source); this provides a
TeX box containing the encapsulated PostScript insert specified by
the file MyFigure. \AffixLabels{...} provides the grid (supposing
\ShowGrid is present) and later, once you have specified labels
using the grid, it will "tack on" the labels.

     The grid is a sort of (usually elongated) checkerboard of
ten rows and ten columns and its (internal) partitions are by
default numbered  .1, ... ,.9  both horizontally (X-coordinate
running left to right) and vertically (Y-coordinate running bottom
to top).  Thus the points enclosed by the grid correspond to the
points of the unit square in the cartesian "X-Y" plane, the lower
left corner corresponding to the origin (0,0).  By extrapolation,
the full page corresponds to a larger rectangle in the plane.

     These coordinates serve to position labels as follows.
Before the \AffixLabels{...} command type label specifications:

  \SetLabels
   (<X-coordinate>*<Y-coordinate>) <first label> \\
   .
   .
   .
   (<X-coordinate>*<Y-coordinate>)  <last label> \\
  \endSetLabels

Each row specifies one label and is terminated by \\.  In each
row, the position indicator comes first; it is written as a
standard cartesian point except that the X- and Y- coordinates
are separated by * rather than a comma because TeX allows a
comma as decimal point. There are no dimension units to specify
as the unit is the grid itself.

     By default, this cartesian point specifies where the middle
of the baseline of the label will be located.  However if you precede
the point by \L [or \R] the left [or right] edge of the baseline will
be located there. Similarly you may also precede the point by \T, \E,
or \B to vertically align the top equator or bottom of the label box
at the specified point.  This gives nine standard positions of
the label with respect to the insertion point --- corresponding to
the eight principle points of the compas and the center

                     \L\T     \T      \R\T

                     \L\E     \E      \R\E

                     \L\B     \B      \R\B

But this neglects the default "baseline" level of TeX,
giving potentially three more positions

                     \L    <no tag>   \R

For text, the baseline level is often the preferred. Its relation to
the others is variable. It will often coincide with the bottom level,
as happens for "X".  But it is often distinct, as for "g", in which
case you have in all 12 distinct positions rather than 9.

     It is convenient to think of this specification of label
position as attaching the label by a thumb-tack to the coordinate
grid. There are up to twelve positions of the thumb-tack on the
label, while the position of the thumb-tack on the coordinate grid is
arbitrary.  Normally, one choses the position of the thumb-tack on
the label to be the one that is the closest to the item being
labeled.  There are good reasons for this "rule of thumb":

   (a)  It facilitates correct positioning at first try.

   (b)  If the scale of the figure must be altered after labels
have been affixed, the labels have a good chance of remaining well
positioned.

   (c)  The visible grid need not extend beyond the "bounding box"
for the figure, because the best preferred position is always
(at least almost) within the bounding box .

The second reason is particularly important. Indeed it often
happens that scale has to be altered after labelling begins, in
order to either provide space for the labels, or to adjust
proportions between the labels and the figure.  (The size of labels
is unaffected by scaling.)

     Here is an artificial but self-contained test which uses
TeX rules to make a graphics object.

TEST

    Do not skip this!

\input labelfig

 \def\FrameIt#1{\hbox{\vrule$\vcenter {\hrule\kern3pt%
             \hbox {\kern3pt #1\kern3pt}%
               \kern3pt\hrule}$\relax\vrule}}

 \def\Caption#1#2{\FrameIt{%
       \vtop {\hsize=#1\relax \parindent=0pt
         \leftskip=0pt \rightskip=0pt plus15pt
         \parfillskip=0pt
         \lineskip=1pt\baselineskip=0pt
         #2}}}

 \def\FirstQuadrant{\hbox to 100pt{\vrule\vbox to 100pt{%
        \hbox to 100pt{\hfil}\vfil\hrule}\hss}}

  \SetLabels
    \R(.5*.2) $\zeta\,\cdot$\\
    (.9*-.10) $\xi$\\
    \R(-.03*.9) $\eta$\\
    \T(.5*.9) \Caption{70pt}{%
          \it The norm of
          $g(\xi+i\eta)$ is indicated on
          contours of this invisible surface.}\\
  \endSetLabels

  \AffixLabels{\FirstQuadrant}

  \end

  Note that the coordinates to use for labels are indicated on the
edges of the grid (when visible) corresponding to the conventional
x- and y- axes of the Cartesian plane. By default the grid is
1-by-1. However, by the command \Edges{100}, you can change this
to 100-by-100 and many users find this alternative most
convenient. Place the command \Edges{...} in your style file (or
header) since its effect is is global. Other possible edge values
are 10 and 1000.

  If you use the command \Edges{...} at all, do so with care.  For
if you accidentally delete an \Edges{...} command your labels will
abruptly be badly misplaced and may logically but mysteriously
generate "dimension too big" errors under TeX and "off page" errors
under your driver.  

  You can dictate the edgescale for an individual figure by giving
the scale in brackets immediately after \SetLabels.  Thus, to
import into an article using say \Edge{100} a figure labelled using
another edgescale, say the original 1-by-1 default, you can use
\SetLabels[1]...\endSetLabels.

GETTING IT DOWN PAT

     Complicated labeling deserves the same respect as
complicated mathematics.  Do not expect it to come out perfect the
first time!  What is needed in either case is a mechanism to
repeatedly typeset troublesome pieces.

     One mechanism is always available.  One does complicated
labelling in a separate "test" file involving just the figure being
labelled;  a texpert will know how to \dump TeX's current state as
a temporary format that restarts rapidly at each retry.  Usually,
one then pastes the completed labelled figure back into the main
TeX file, but, of course, one can also \input it as an auxiliary
file.

     If you do not have a TeXpert at handy, here is a first
approximation to an efficient setup. By deletions reduce a copy
of your article to just a few lines before and after the figure.
Now label the figure, and finally, copy and paste the labelled
figure to the original article. Then copy the next figure to label
into this testbed and repeat. The TeXpert can improve the  speed
at which TeX starts up, by compiling a format specifically for
your article; just one caution: best NOT include in the format
ephemeral details of setup like \Set<mydriver>ArtSpecials (from
boxedeps.tex because this reads  figure dimensions which you may
change during your work session.

     An improved mechanism to repeatedly typeset troublesome
pieces is now available on the Macintosh; it is called LinoTeX;
see the same ftp sources.  It could be set up on many types
of computer.

     Before using labelfig.tex to attach labels to a graphics
object inserted using boxedeps.tex or BoxedArt.tex, make it a
firm rule to carefully adjust the bounding box using the trimming
commands of these packages, and also at least tentatively scale
and position the object. Beware of changing the grid inadvertently
after the labels have been positioned.  For example, correcting
the bounding box of a PostScript graphics object can foul up the
labels by changing the coordinate grid to which the labels are
attached. This is particularly true for the trimming  commands of
boxedeps.tex and BoxedArt.tex. However, as noted already, change
of scale is much less disruptive, and modest adjustments should be
well tolerated.

     Sometimes the labels protrude so far from the bounding box
of a figure that the figure has to be repositioned.  Best do this
by ad hoc spacing, say using \hglue and \vglue; altering the
bounding box would create a vicious circle.

     Remember that you are responsible for preventing labels
from overlapping. You are responsible for all label typography
including size and style. A label is really just about anything
that can be put in a TeX box. Note that spaces at the beginning
and end of labels will normally be suppressed; if you really want
them you must protect them with TeX braces.

     This package temporarily sets the \mathsurround parameter
of TeX to zero  while the labels are being affixed. This is done
because nonzero \mathsurround space would influence the position
of left and right aligned labels; then, when a texpert or printer
modifies mathsurround, diagram labeling might be disastrously
altered. There is a small price to pay involving labels that are
formatted as caption boxes including mathematics: you  may want or
need to specify an explicit mathsurround space within the caption
box; it will not influence anything outside.

     Those hostile to the use of * as separator between
the X and Y coordinates of label insertion points, are free to
impose another using \SetXYSeparator{<the new separator>}.  
Americans may prefer "," to "*" since they never use a 
comma as a decimal point; on the other hand, * may be more visible.

APPENDIX (I)  MERGING labelfig.tex LABELS INTO AN EPSF GRAPHICS OBJECT.

     As promised in the introduction, here is a recipe useful for
publishers. It works at least on Macintosh and at least for vectorized
graphics and Adobe type1 fonts.  (There is surely a similar recipe for
PCs under MSWindows.)

 (a)  Use boxedeps.tex utility to integrate the figure given by the eps
file, "x.eps" say, with a visible frame around it.  See
\ShowDisplacementBoxes command in boxedeps.tex.  To get precise results
automatically it is important to use the \Trim... commands of
boxedeps.tex making the "DisplacementBox" neatly fit the figure.

 (b)  Use the TeX printer driver and LaserWriter (versions >= 8.1.1) to
export to an EPSF the DVI page containing the integrated, labelled
figure. You now have an EPS file  "xx.eps"  that contains too much, and at
the wrong scale, and at wrong position.

 (c)  Convert the EPSF to an Adode Illustrator format EPSF using
the shareware utility called epsConvert by Sam Weiss
1993-- (currently $25).

 (d)  In Illustrator (or a compatible program), group the labels and the
"DisplacementBox"; copy them to the clipboard and paste them into "x.ps".
This step requires that all the label fonts be "visible to the Macintosh.

 (e)  Translate and scale the pasted group consisting of the labels plus
the "DisplacementBox" so as to make the "DisplacementBox" the bounding
box of (labelless) figure represented by "x.eps".  At this point the
labels will be correctly placed on the figure "x.eps".

 (f)  Ungroup and delete the "DisplacementBox".  The result is the
desired single EPS file, "x+.eps" say, It contains the original figure
plus its labels.  

     Using grouping and ungrouping appropriately in "x+.eps", a
publisher's staff can very efficiently improve label positions etc.

APPENDIX II)  SOME EXOTIC APPLICATIONS

     The grid of labelfig.tex is analogous to a light-table in
classical page makeup with wax or latex glue.  In principle, you
can use it to compose any page from its indivisible parts.  This
even has some of the artisanal charm of classical paste-up
provided you have a fast screen preview to make the process
"interactive".

     In practice labelfig.tex is a tool for nonstandard jobs.
Here are a few going beyond the labelling already discussed.

(I)  GRAPHICS INTEGRATION.

     This is accomplished by treating the imported graphics
objects as labels.  The underlying graphics object is then
typically an empty  \vbox to <dimension>{\vfill} in a TeX
\midinsert...\endinsert construction.  A label line
might be of the form

   (.1*.1) \\

The exact form of the special command varies from driver to
driver.  However, in the case of encapsulated PostScript graphics
(EPSF norm), by relying on boxedeps.tex, one can have the
following standard syntax (independant of driver  (see
boxedeps.doc for details.
  
  (.1*.1) \BoxedEPSF{MyFigure scaled <scale in mils>}\\

This may be slow since it requires TeX to read the PostScript
file to read bounding box using many complex macros.  So you
may want to try

  (.1*.1) \EPSFSpecial{MyFigure}{<scale in mils>}\\

which is fast and driver independant, but it squashes the
bounding box, normally to its lower left corner.

     Similarly for graphics of the Macintosh PICT norm ---
using BoxedArt.tex (same sources) in place of boxedeps.tex.

     This approach to integration is to be recommended when
one is assembling a composite graphics object.

 (II)  COMMUTATIVE DIAGRAM ENHANCEMENT

     Commutative diagrams or arrays of mathematical objects
connected by arrows of various sorts are common in mathematics.
The mathematical objects require the use of TeX.  Recently TeX
acquired a good collection of arrows of all slopes --- that of
LamSTeX --- plus pwerful macros to build the diagrams.

     However, even the LamSTeX collection is often
inadequate; it lacks for example double shafted arrows, dotted
arrows and curved arrows. Fortunately it is possible to produce
such arrows on an individual basis using sophisticated graphics
programs such as Illustrator and AldusFreehand (both serving
the EPSF norm) or using Metafont (with its public domain norm).
Since the creation of each new arrow is a work of love, you
probably want to limit the number of arrows by using LamSTeX
for most arrows. The 40K commutative diagram module of LamSTeX
has been adapted to work with AmSTeX and a copy may be posted
with LabelFig and related files. Unfortunately no one has yet
offered a version that works with Plain TeX or LaTeX.

       Suffice it here to say that when the exotic arrow has
been somehow imported into TeX, labelfig.tex treats it as a
label that one affixes to the commutative diagram.  Two other
steps will be treated in separate notes, namely the matter of
extracting the dimension specifications for the arrow and the
construction of the arrow --- for these steps are far from
unique and often depend intimately on your computer environment. 
Notes for the Macintosh-Textures-Illustrator combination are
found in the file ExoticArrows.doc.

 (III) NESTING 

Ingenuity pays off in exploiting labelfig.tex. One can
mix graphics and typography quite freely.  labelfig.tex is good
for freeform or overlapping arrangements, while boxedeps.tex (or
BoxedArt.tex) is best for regimented non-overlapping
arrangements --- and the two can be combined.

     The default behavior of labelfig.tex is not ideal 
for nesting objects, because to prevent trouble for beginners
the register for labels is globally cleared when \AffixLabels
concludes.  But there are switches available

      \LabelsGlobal      \LabelsLocal

which change this.  To understand this, extend the above test 
by something like:

 \LabelsLocal

 \SetLabels
    (.5*.5) AAA\\
 \endSetLabels

 {
 \SetLabels
    (.5*.5) ZZZ\\
 \endSetLabels
   \AffixLabels{\FirstQuadrant}
 }

   \AffixLabels{\FirstQuadrant}

     There are however potential pitfalls.  Neither
labelfig.tex nor boxedeps.tex has been tested under extreme
conditions. Problems may occur if their procedures are
indiscriminately nested. For boxedeps.tex (not labelfig.tex)
there is a precise cause for worry, namely many of its
variables are "global", which means that TeX braces will not
provide the protection one might expect.

COMMAND SUMMARY FOR labelfig.tex

  Here [...] means optional (one or zero)
       [...]* means any number of such constructs

  \SetLabels
    [[<P>](<X><Sep><Y>) <label> \\]*
  \endSetLabels
  \ShowGrid  
  \AffixLabels{<the figure>}

   --- <P> is tack position, one of eleven or empty
              order irrelevant

                   \L\T      \T      \R\T

                   \L\E      \E      \R\E

                     \L               \R

                   \L\B      \B      \R\B

   --- (<X><Sep><Y>) insertion point;
  <Sep> is separator, = * by default;
  \SetXYSeparator{<Sep>} changes it.
   <X> and <Y> are real numbers

  --- <label> a label to attach 

  --- <the figure> the figure to label 

  \GlobalLabels (default)     
  \LocalLabels  setting for nested constructs.

 \Grids makes ALL grids appear; \HideGrid then makes just next disappear.
 \noGrids returns to default.  The commands are always global.

 \GridLineWidth{<dimension>} adjusts width of grid lines. Default is very
small, to give "hairline" effect. If your grid lines are missing try
setting \GridLineWidth{1pt}.

 \Edges#1 globally changes the edge size of all grids to the numerical 
value #1, which must be 1, 10, 100, or 1000.  The default is 1.

VERSION HISTORY.
 --- Jan 1993: \Edges#1 and [??] option after \SetLabels
 --- July 1992: \Grids, \noGrids, \HideGrid;
       Gridlines become hairlines; \GridLineWidth{<dimension>}.
 --- Oct 1991, Jan 1992: \SetXYSeparator{<Sep>},  \LabelsGlobal,
       \LabelsLocal.
 --- July 1991: first release

Address for bugs and other feedback:

        Raymond S\'eroul
        IREM and Lab. de Typographie Informatise
        Univ. Rene Descartes
        Strasbourg

    Tel 33-88-41-63-45
    Email:  A18645@FRCCSC21.BITNET

        Laurent Siebenmann
        Mathematique, Bat. 425,
        Univ de Paris-Sud,
        91405-Orsay,
        France

    Tel 33-1-6941-7949; 
    Email: lcs@topo.math.u-psud.fr